\documentstyle[preprint,aps]{revtex}

\begin{document}
\title{Ultrafast collective dynamics of short-range charge/orbital ordering
fluctuations in colossal magnetoresistive oxides }
\author{Y. H. Ren$^{1}$, Y. F. Hu$^{2}$, Qi Li$^{2}$, C. S. Hong$^{3\dagger }
$, N. H. Hur$^{3}$, and G. L\"{u}pke$^{1}$}
\address{$^{1}$Applied Science, the College of William \& Mary,\\
Williamsburg, VA 23187\\
$^{2}$Department of Physics, Pennsylvania State University, University Park,%
\\
PA 16802\\
$^{3}$Center for CMR Materials, Korea Research Institute of Standards and\\
Science, P. O. Box 102 Yusong, Daejeon, 305-600, Republic of Korea}
\date{\today }
\maketitle

\begin{abstract}
{\bf The ``colossal magnetoresistive'' (CMR) manganites are highly
correlated systems with a strong coupling between spin, charge, orbital, and
lattice degrees of freedom, which leads to complex phase diagrams and to the
coexistence of various forms of ordering. For example, nanoscale
charge/orbital ordering (CO) fluctuations appear to cooperate with
Jahn-Teller (JT) distortions of the MnO}$_{{\bf 6}}${\bf \ octahedra\ in CMR
manganites and compete with the electron itinerancy favored by double
exchange.} {\bf However, access to the ordered dynamical state has been
challenging, mostly due to intrinsic experimental difficulties in measuring
fast short-range correlations. Here, we report on a strongly damped
low-energy collective mode originating from fast short-range CO fluctuations
in La}$_{0.67}${\bf Ca}$_{0.33}${\bf MnO}$_{3}$\ {\bf (LCMO) single crystal
and thin films. We elucidate the collective mode in terms of its dispersion
relation and dependence on average A-site ion radius, r}$_{{\bf A}}${\bf ,
and hole-doping concentration. Our results show for the first time that
dynamical short-range CO correlations in CMR manganites can be detected with
high\ momentum resolution by coherent ultrafast optical techniques.}
\end{abstract}

Charge/orbital ordering (CO) is a common characteristic of transition metal
oxides with perovskite structure \cite%
{MillisNature1998,CoeyAdvPhys1999,ImadaRMP1998,DessauScience2001}. Recent
neutron scattering experiments showed that fast short-range correlation of
the so-called CE-type charge ordering exists in the colossal
magneto-resistive (CMR) manganite, LCMO, above the Curie temperature, $T_{C}$
\cite{LynnPRL1996,TeresaNature1997}. The coupled dynamics of fluctuating CO
phases and Jahn-Teller (JT) distortions of the MnO$_{6}$\ octahedra are
found to be crucial for the metal-insulator (MI) phase transition and the
CMR effect \cite{DaiPRL2000,AdamsPRL2000}. Therefore, it is fundamentally
important to elucidate the dynamics of short-range charge/orbital stripes in
CMR materials.

Ultrafast optical techniques have provided significant insight into coupled
electron, lattice and spin dynamics in metals \cite{Beaurepaire PRL,Koopmans
PRL}, and more recently, transition metal oxides \cite{Dodge PRL,Kise
PRL,RastPRB2001,Ren JAP,RenPRB2001}. In this work, we investigate for the
first time the collective dynamics of short-range CO fluctuations in CMR
manganites by time-resolved\ optical spectroscopy. The period of the CO
modulations show strong dependence on hole doping concentration and average
A-site ion radius, which is explained by different correlation length and
strength in these materials. We obtain a coherence length of $\sim $250 nm
for LCMO\ and $>>$2.5 $\mu $m for LaMnO$_{3}$ (LMO). The overdamped behavior
in LCMO is due to inhomogeneous dephasing, whereas in LMO, the damping rate
is very small because of the static and long-range correlation. Our results
show that {\it the time-resolved optical technique provides
momentum-resolved spectroscopic information }on the low-energy dynamics of
short-range CO fluctuations in colossal magneto-resistive materials. This
important new information is relevant for elucidation of collective
transport phenomena in strongly correlated electron systems.

In the time-resolved optical experiments, the change of reflectivity ($%
\Delta R$) induced by the pump beam is measured as a function of time delay
for different wavelengths of the probe beam ($\lambda _{probe}=400$ nm - 2.5 
$\mu $m) in both LCMO single crystal and thin films. Figure 1 shows typical
time evolution of $\Delta R$ for LCMO thin film and LMO single crystal as a
function of time delay $\Delta t$ between the pump and probe pulses. The
data are taken at 285 K and 35 K with pump and probe wavelength of 800 nm,
respectively. After the initial laser pulse excitation, the decay of $\Delta
R$ clearly shows an oscillatory component on top of a multi-exponential
decay.

Here, we discuss only the slow oscillatory component observed in the trace
of $\Delta R$ from LCMO and LMO (Fig. 1). Figure 2 displays the coherent
overdamped oscillations of $\Delta R$ from LCMO thin film without the
exponentially decaying part as a function of probe wavelength in the range
400 nm to 2300 nm. The oscillations are heavily damped and strongly
dispersive. Only one, at most two oscillations, are observed in the whole
wavelength range. The frequency of the oscillations decreases from 73.83 GHz
at 400 nm to 21.8 GHz at 2300 nm.

Figure 3 shows the dispersion relation of the slow oscillation of $\Delta R$
from LCMO thin film at 285 K. The wave number is given by $q=2n/\lambda $,
where $\lambda $ is the probe wavelength and $n$ is the refractive index of
LCMO \cite{KIMPRL}. The frequency of the coherent oscillations increases
proportional to the wave number of the excited mode. The slope gives a phase
velocity for the\ collective mode of $c_{\phi }=$ $7.1\pm 0.3\times 10^{3}$
m/s along the $c$-axis.

The coherent oscillations of $\Delta R$\ show strong dependence on hole
doping concentration and average A-site ion radius, $r_{A}$. The former is
clearly revealed in $\Delta R$ from LMO (Fig. 1). Similar to the LCMO
sample, $\Delta R$ also exhibits a slow oscillatory component on top of a
bi-exponential decay in the parent compound LMO. However, the 800-nm data
from LMO reveal a different modulation period of $\sim $ 45 ps, in contrast
to $\sim $ 27.6 ps observed in LCMO (Fig. 1). The oscillations also show a
strong dependence on A-site substitution (average A-site ion radius, $r_{A}$%
). The modulation period for PCMO single crystal at 800 nm is $\sim $ 40 ps %
\cite{FiebigAPB2000}, which is much longer than in LCMO.

These observations strongly suggest that the coherent oscillations originate
from a low-energy collective mode in LCMO. The fact that the period of the
oscillations is longer in the parent compound LMO and PCMO, which has a
smaller average ionic radius than LCMO, can be explained by different
correlation length and strength in these materials. As compared to LCMO,
where the charge/orbital correlations are short-range in nature due to the
presence of ferromagnetic phases, the charge/orbital coupling is stronger
and long-range in both LMO and the smaller bandwidth manganite PCMO. The
case of LMO, with no double-exchange carriers, is a typical example in which
the orbital ordering is enhanced by the increase of Jahn-Teller interaction.
The stronger charge/orbital correlation in LMO results in a heavier
effective mass than in LCMO. The phase velocity of the collective modes is
determined by the effective mass and the Fermi velocity, $v_{F}$ \cite{Fermi}%
:

\[
c_{\phi }=(\frac{m}{m^{\ast }})^{1/2}v_{F} 
\]
which is therefore smaller in LMO than in LCMO. We expect a longer
oscillation period in LMO since the modulation period is inversely
proportional to the phase velocity of the material. Similarly, the ground
state of PCMO exhibits charge-localizing real space ordering of Mn$^{3+}$
and Mn$^{4+}$ ions that occurs at \ $\sim $220 K. The charge/orbital
correlation in PCMO is enhanced by the increase of orthorhombic distortion
of GdFeO$_{3}$-type \cite{CoeyPRL1995}. Thus, a longer oscillation period is
expected in PCMO as well.

Another particularly intriguing result is the strong doping concentration
dependence of the damping rate of the coherent oscillations in LCMO. As
shown in Fig. 1, the coherent oscillations in LMO persist for at least 600
ps ($\sim $15 periods) with very little damping whereas at most two periods
of the collective mode can be observed in LCMO. Since the optical properties
of LCMO and LMO are very similar at photon energies in the near infrared to
visible range \cite{QuijadaPRB1998,QuijadaPRB2001}, the stronger damping
observed in LCMO must be directly related to the damping of the collective
mode. The results shown in Fig. 1 allow us to extract the coherence length

\[
\lambda =\tau c_{\phi } 
\]
for LCMO and LMO, which depends essentially on the damping time $\tau $ of
the collective mode. We obtain a coherence length of $\sim $250 nm for LCMO\
and $>>$2.5 $\mu $m for LMO. The overdamped behavior in LCMO may result from
inhomogeneous spatial distribution (inhomogeneous dephasing) and the
dynamical nature of the short-range charge/orbital ordering phases. The
coherence length for LCMO is of the same order of magnitude as the
charge/orbital correlation length of the CE-type phase in perovskite
manganites with a characteristic length scale of stripe domains on the order
of 100 lattice spacings \cite{NelsonPRB2002}. In contrast, the damping rate
is very small for LMO, because the orbital domains in LMO are static and the
correlation is long-range. There is also a very limited phase space
available for scattering into other modes in the wavelength range
investigated.

The inset of Fig. 1 illustrates the optically excited collective modes in
LCMO\ and LMO. The photo-generated charge/orbital ordering modulations cause
complex rotations of the MnO$_{6}$ octahedra, which oscillate along the
modulation direction. The collective\ modes propagate from the surface in
the crystallographic c-direction and approach zero with increasing distance
from the origin due to the finite range of correlations.

Finally, we present a simple physical model that accounts correctly for the
excitation and detection of the coherent charge/orbital-ordering
oscillations. The number of photons per pulse and unit volume absorbed in
the sample is $\sim $10$^{20}$ photons/cm$^{3}$, comparable to the
charge-carrier density ($\sim $10$^{20}$-10$^{21\text{ }}$holes/cm$^{3}$) in
LCMO; hence, one expects significant electron excitation during ultrashort
pump pulse illumination. This can lead to coherent excitation of collective
modes \cite{CM}. We use a tunable optical probe pulse to detect the various
frequency components of the photo-generated collective modes. In the
back-scattering geometry (Fig. 4), part of the probe pulse is reflected by
the wave front of the excited collective modes and the remainder at the
surface of the sample. These reflections interfere constructively or
destructively depending on the position and time of the charge density
modulation. Further, the momentum selection rule for back scattering is $%
q_{i}+q_{f}=q\cos \theta $, where $q_{i}$ and$\ q_{f}$ are the wave vectors
of the incident and scattered probe beam in the material,$\ $and $\theta $
is the probe beam incident angle ($\theta \sim 0^{\circ }$). Therefore, for
a given probe wavelength phase-matching occurs exclusively for a single
collective mode wavevector. This process causes the probe signal to
oscillate with time delay relative to the pump\ pulse. The new scheme
provides momentum-resolved spectroscopic information of collective modes in
solids.

In summary, we investigated fast short-range CO fluctuations in LCMO by
time-resolved optical spectroscopy. The amplitude, period and\ damping of
the CO modulations show strong dependence on hole doping and average A-site
ion radius. The results reported here have great significance for
understanding the competition/cooperation behavior and nanoscale phase
separation in colossal magnetoresistive manganites. Our scheme represents a
new approach to investigate low-energy collective dynamics in strongly
correlated electron systems with high-momentum resolution.

\bigskip

\bigskip

\bigskip

\bigskip

{\bf Methods}

LCMO and LMO single crystals and thin films were grown by the floating zone
method and pulsed laser deposition. The samples were characterized by
electrical resistivity and magnetization measurements. The LCMO single
crystal and 400-nm thin film (grown on NdGaO$_{3}$ (110) substrate) have a
Curie temperature $T_{C}$ = 225 K and 260 K, respectively. The LMO single
crystal shows a Neel temperature $T_{N}$\ = 145 K. For the transient
reflectivity measurements the samples were mounted in an optical cryostat.
The laser system consists of a Ti:sapphire regenerative amplifier (Spitfire,
Spectra-Physics) and an optical parametric amplifier (OPA-800C,
Spectra-Physics)\ delivering 100-fs short pulses at a 1-kHz repetition rate
tunable from 600 nm to 10 $\mu $m. A two-color pump-probe setup is employed
with the pump beam power $<$ 6 mW and the probe beam power $<$ 1 mW. The
unfocused pump beam, spot-diameter $\sim $2 mm, and the time-delayed probe
beam are overlapped on the sample with their polarization perpendicular to
each other. The reflected probe beam is detected with a photodiode detector.
A SR250 gated integrator \& boxcar averager, and a lock-in amplifier are
used to measure the transient reflectivity change $\Delta R$ of the probe
beam.

\bigskip

$^{\dagger }$ Current address for C. S. Hong, Department of Chemistry, Korea
university, Anam-dong, Sungbuk-ku, Seoul 136-701, Republic of Korea

\bigskip

\bigskip

\bigskip

{\bf Acknowledgements}

This work is supported in part by NSF through grants: DMR-0137322,
IMR-0114124 (CWM), and DMR-9876266 (PSU), and the Petroleum Research Fund
(PSU).

Correspondence and requests for materials should be addressed to G. L.

\bigskip

\bigskip

\bigskip

{\bf Competing financial interests}

The authors declare that they have no competing financial interests.

\bigskip

\bigskip

\bigskip

\bigskip

\bigskip

\bigskip

{\bf Figure captions}

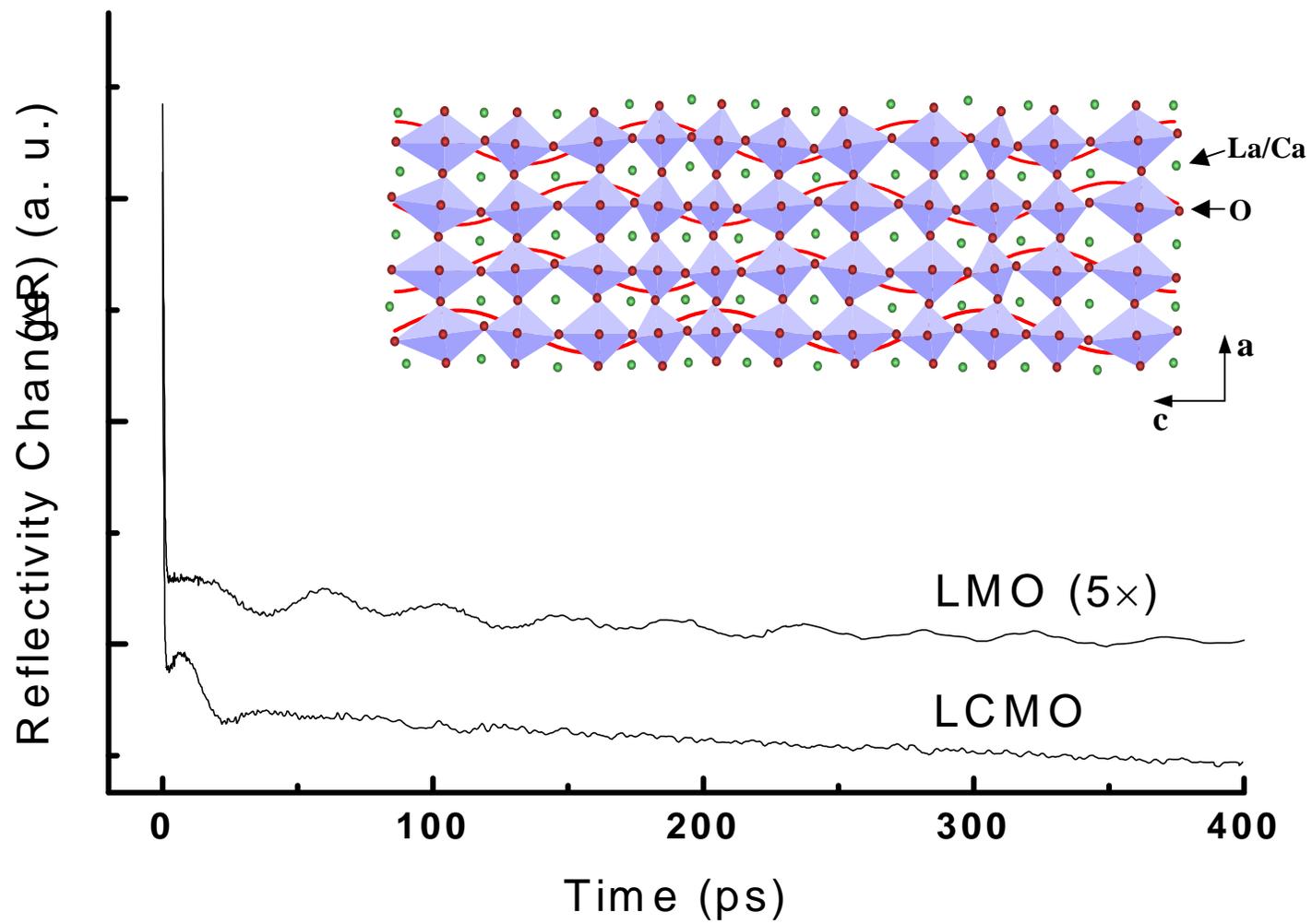
Figure 1. Time evolution of reflectivity change, $\Delta R$, for LCMO thin
film at 285 K and LMO single crystal at 35 K. The transient reflectivity
clearly reveals coherent oscillations on top of a multi-exponential decay.
The inset shows a schematic illustration of optically excited collective
modes in LCMO\ and LMO. The photo-generated charge/orbital ordering
modulations cause complex rotations of the MnO$_{6}$ octahedra, which
oscillate along the modulation direction and approach zero with increasing
distance from the origin due to the finite range of correlations.

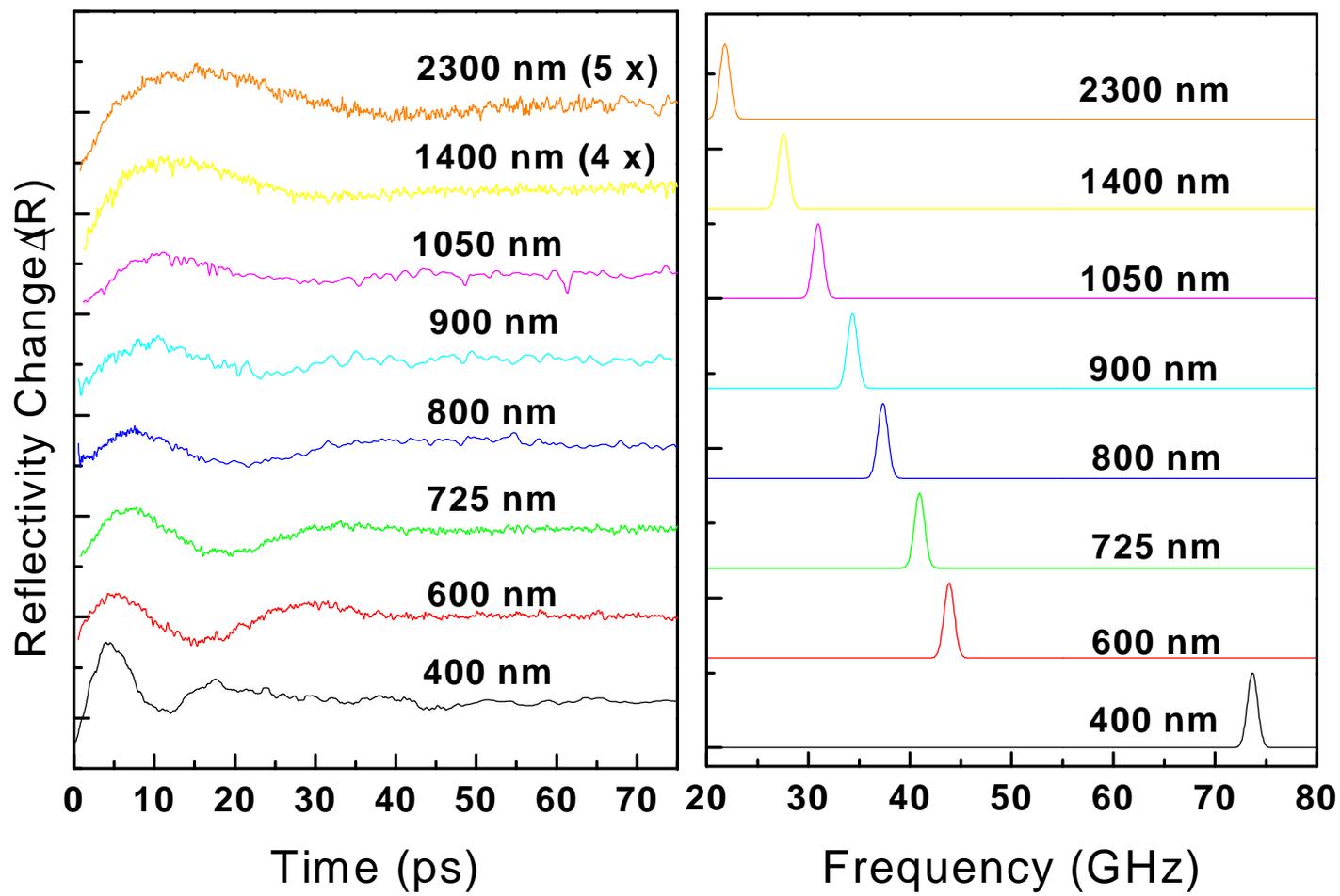
Figure 2. a) Time evolution of $\Delta R$ from LCMO thin film for probe
wavelengths from ultraviolet (400 nm) to mid-infrared (2300 nm); b) Fourier
transforms of $\Delta R$ traces shown in a).

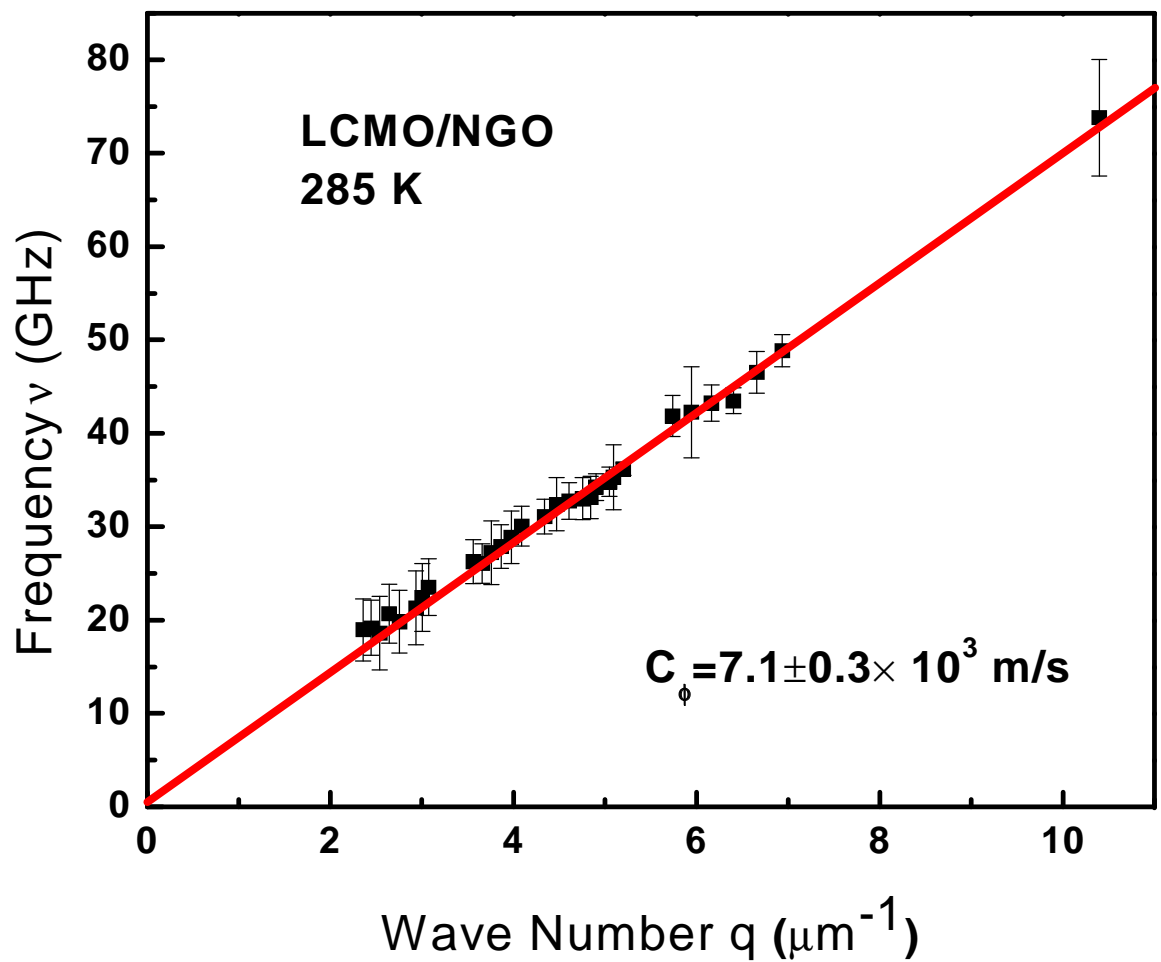
Figure 3. Collective mode dispersion relation of LCMO thin film with
thickness 400 nm grown on NdGaO$_{3}$ (110) substrate at 285 K. Solid
squares: the experimental data. Solid line: the predicted dependence $\nu
=2nc_{\phi }\cos \theta /\lambda $, with $\lambda $ the probe wavelength, $n$
the refractive index of LCMO, and $\theta $ the probe beam incident angle ($%
\theta \sim 0^{\circ }$).

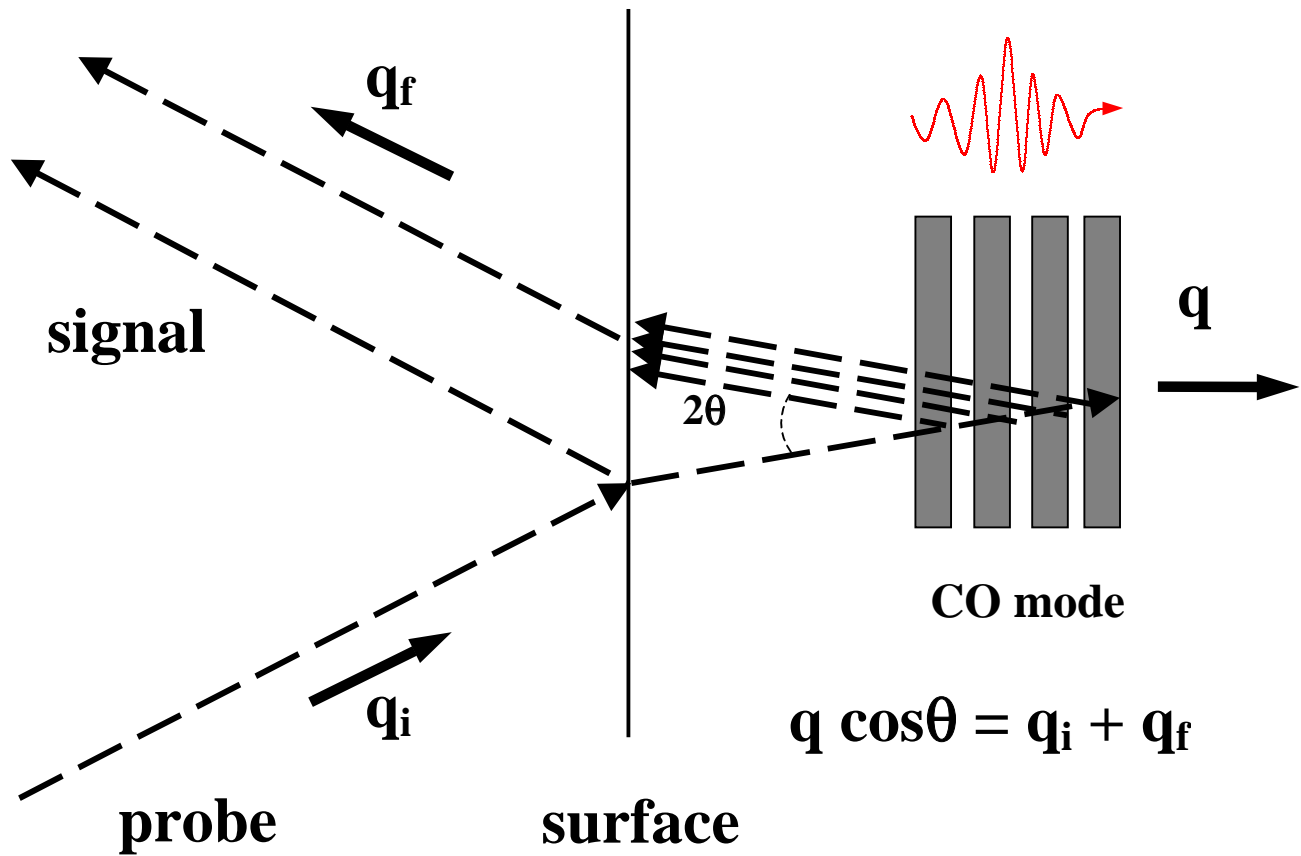
Figure 4. Schematic of photo-induced coherent CO mode excitation and
detection mechanism. The\ strong absorption\ of ultrashort pump pulses
generates broad-band coherent CO oscillations. The modes propagate into the
sample and are detected in the back-scattering geometry by tunable optical
probe pulses. Momentum selection rule for back scattering is $%
q_{i}+q_{f}=q\cos \theta $, with $q_{i}$ and$\ q_{f}$ the wave vectors of
the incident and scattered probe beam in the material.


\begin{references}
\bibitem{MillisNature1998} Millis, A. J., Lattice effects in
magnetoresistive manganese perovskites, {\it Nature} {\bf 392}, 147 (1998).

\bibitem{CoeyAdvPhys1999} Coey, J. M. D., Viret, M., and Molnar, S. von,
Mixed-valence manganites, {\it Adv. Phys}. {\bf 48}, 167 (1999).

\bibitem{ImadaRMP1998} Imada, M. I., Fujimori, A., Tokura, Y.,
Metal-insulator transitions, {\it Rev. Mod. Phys}. {\bf 70}, 1039 (1998).

\bibitem{DessauScience2001} Chuang, Y. -D., Gromko, A. D., Dessau, D. S.,
Kimura, T., and Tokura, Y., Fermi surface nesting and nanoscale fluctuating
charge/orbital ordering in colossal magnetoresistive oxides, {\it Science} 
{\bf 292}, 1509 (2001).112.

\bibitem{LynnPRL1996} Lynn, J. W., {\it et al.} Unconventional ferromagnetic
transition in La$_{1-x}$Ca$_{x}$MnO$_{3}$, {\it Phys. Rev. Lett}. {\bf 76},
4046 (1996).

\bibitem{TeresaNature1997} Teresa, J. M. De {\it et al.}, Evidence for
magnetic polarons in the magnetoresistive perovskites, {\it Nature} {\bf 386}%
, 256 (1997).

\bibitem{DaiPRL2000} Dai, Pengcheng, {\it et al.} Short-range polaron
correlations in the ferromagnetic La$_{1-x}$Ca$_{x}$MnO$_{3}$, {\it Phys.
Rev. Lett}. {\bf 85}, 2553 (2000).

\bibitem{AdamsPRL2000} Adams, C. P., Lynn, J. W., Mukovskii, Y. M., Arsenov,
A. A., and Shulyatev, D. A., Charge ordering and polaron formation in the
magnetoresistive oxide La$_{0.7}$Ca$_{0.3}$MnO$_{3}$, {\it Phys. Rev. Lett}. 
{\bf 85}, 3954 (2000).

\bibitem{Beaurepaire PRL} Beaurepaire, E., Merle, J.-C., Daunois, A., and
Bigot, J.-Y., Ultrafast spin dynamics in ferromagnetic nickel, {\it Phys.
Rev. Lett. }{\bf 76}, 4250 (1996).

\bibitem{Koopmans PRL} Koopmans, B., van Kampen, M., Kohlhepp, J. T., and de
Jonge, W. J. M., Ultrafast magneto-optics in nickel: magnetism or optics?, 
{\it Phys. Rev. Lett. }{\bf 85}, 844 (2000).

\bibitem{Dodge PRL} Dodge, J. S., {\it et al.}, Time-resolved optical
observation of spin-wave dynamics, {\it Phys. Rev. Lett.} {\bf 83}, 4650
(1999).

\bibitem{Kise PRL} Kise, T., {\it et al.}, Ultrafast spin dynamics and
critical behavior in half-metallic ferromagnet: Sr$_{2}$FeMoO$_{6}$, {\it %
Phys. Rev. Lett.} {\bf 85}, 1986 (2000).

\bibitem{RastPRB2001} Rast, S., {\it et al.},{\bf \ }Evidence for two
coupled subsystems in the superconducting state of La$_{2-x}$Sr$_{x}$CuO$%
_{4} $, {\it Phys. Rev. B} {\bf 64}, 214505 (2001).

\bibitem{Ren JAP} Ren, Y. H., Zhao, H. B., L\"{u}pke, G., Hu, Y. F., Li, Qi,
Strain-dependent spin dynamics in Nd$_{0.67}$Sr$_{0.33}$MnO$_{3}$ near the
metal--insulator transition, {\it J. Appl. Phys}. {\bf 91}, 7514 (2002).

\bibitem{RenPRB2001} Ren, Y. H., {\it et al.}, Observation of strongly
damped GHz phonon-polariton oscillations in La$_{0.67}$Ca$_{0.33}$MnO$_{3}$, 
{\it Phys. Rev. B,} {\bf 64}, 144401 (2001).

\bibitem{KIMPRL} Kim, K. H., Jung, J. H., and Noh, T. W., Polaron absorption
in a perovskite manganite La$_{0.7}$Ca$_{0.3}$MnO$_{3}$, {\it Phys. Rev. Lett%
}. {\bf 81}, 1517 (1998).

\bibitem{FiebigAPB2000} Fiebig, M., Miyano, K., Tomioka, Y., and Tokura, Y.,
Sub-picosecond photo-induced melting of a charge-ordered state in a
perovskite manganite, {\it Appl. Phys. B} {\bf 71}, 211 (2000).

\bibitem{Fermi} The Fermi velocity $v_{F}$ is similar in LCMO and LMO. See
e.g. Singh, D. J., Pickett, W. E., Pseudogaps, Jahn-Teller distortions, and
magnetic order in manganite perovskites, {\it Phys. Rev. B} {\bf 57}, 88
(1998).

\bibitem{CoeyPRL1995} Coey, J. M. D., BViret, M., Ranno, L., and Ounadjela,
K., Electron localization in mixed-valence manganites, {\it Phys. Rev. Lett}%
. {\bf 75}, 3910 (1995).

\bibitem{QuijadaPRB1998} Quijada, M., {\it et al.}, Optical conductivity of
manganites: crossover from Jahn-Teller small polaron to coherent transport
in the ferromagnetic state, {\it Phys. Rev. B} {\bf 58}, 16093 (1998).

\bibitem{QuijadaPRB2001} Quijada, M. A., {\it et al.}, Temperature
dependence of low-lying electronic excitations of LaMnO$_{3}$, {\it Phys.
Rev. B} {\bf 64}, 224426 (2001).

\bibitem{NelsonPRB2002} Nelson, C. S., {\it et al.}, Coherent x-ray
scattering from manganite charge and orbital domains, {\it Phys. Rev. B} 
{\bf 66}, 134412 (2002).

\bibitem{CM} Since the penetration depth, $\xi $, in LCMO is very small ($%
\sim $100 nm), the momentum bandwidth of the excited coherent collective
modes must be large: from $\delta q\delta x>1$ and $\delta x\sim \xi \sim
100 $ nm follows that $\delta q>1\times 10^{5}$ cm$^{-1}$, i.e., larger than
the biggest $q$ probe used. Hence, we expect to generate a broad band of
collective modes, comparable to the entire range of probe wavelengths.
\end{references}
\end{document}